\documentclass [12pt]{article}

\begin{document}
\begin{center}
\textbf{\huge Entropy of Kerr-Newman black hole to all orders in
the Planck length}
\end{center}

\begin{center}

 Zhao
Ren\footnote{e-mail address: zhaoren2969@yahoo.com.cn },Zhang
Li-Chun,  Li Huai-Fan  and Wu Yue-Qin

Department of Physics, Shanxi Datong University, Datong 037009
P.R.China

\end{center}
\vspace{1.0cm}
\begin{center}
\textbf{Abstract}
\end{center}

Using the quantum statistical method, the difficulty of solving
the wave equation on the background of the black hole is avoided.
We directly solve the partition functions of Bose and Fermi field
on the background of an axisymmetric Kerr-Newman black hole using
the new equation of state density motivated by the generalized
uncertainty principle in the quantum gravity. Then near the black
hole horizon, we calculate entropies of Bose and Fermi field
between the black hole horizon surface and the hypersurface with
the same inherent radiation temperature measured by an observer at
an infinite distance. In our results there are not cutoffs and
little mass approximation introduced in the conventional
brick-wall method. The series expansion of the black hole entropy
is obtained. And this series is convergent. It provides a way for
studying the quantum statistical
entropy of a black hole in a non-spherical symmetric spacetime.\vspace{1cm}
\vspace{4cm}\\
\textbf{\large 1. Introduction}

In the early 1970s, Bekenstein,Hawking, Bardeen et al. proposed
that the black hole entropy is proportional to the area of event
horizon [1-3]. Hereafter research on black hole thermodynamics
makes rapid progress. Especially the proof of Hawking radiation
has been effective in stimulating the enthusiasm of scientists for
black hole thermal property [2]. If the gravity at a black hole's
surface $\kappa $ is taken as temperature and the area of horizon
A is taken as the entropy, four laws of thermodynamics in black
hole' s theory can be derived. Hereafter, black hole
thermodynamics has received considerable attention. Especially
researchers pay widespread attention to the statistical origin of
the black hole entropy. Many methods of calculating entropy have
emerged[4-9]. One of them is the well-known brick-wall method[7].
Statistical properties of free scalar fields on background of
various black holes are discussed by this method [10-14]. The
entropy expression with respect to the horizon area is derived. It
is shown that the entropy is directly proportional to its outer
horizon area. For Schwarzschild spacetime [7], when cutoffs
satisfy a proper condition, the entropy can be written as $S =
\raise0.7ex\hbox{${A_H }$} \!\mathord{\left/ {\vphantom {{A_H }
4}}\right.\kern-\nulldelimiterspace}\!\lower0.7ex\hbox{$4$}$. When
cutoffs approach zero, the entropy diverges. 't Hooft thought that
this divergence was caused by the fact the state density would
approach infinite near the horizon. Subsequently, it is found that
the quantum states make a leading contribution to the black hole
entropy near the horizon. Thus, the brick-wall model has been
improved and the thin layer model has been proposed [15-17]. The
thin layer model only considers the quantum states in a thin layer
near the horizon. The infrared cutoff and little mass
approximation in the original brick-wall model are avoided.
However, there still exists ultraviolet cutoff. Recently, It is
found that the generalized uncertainty principle is related to
state density [18-23]. The statistical entropies have been
calculated using the generalized uncertainty principle [24-30].
The simplest way to generalize the uncertainty principle is to
promote it to [28]

\begin{equation}
\label{eq1}
\Delta x\Delta p \ge \frac{1}{2}\left( {1 + \lambda (\Delta p)^2} \right)
\end{equation}
and the correction to state density is

\begin{equation}
\label{eq2}
dn = \frac{d^3xd^3p}{(2\pi )^3(1 + \lambda p^2)^3}
\end{equation}
The statistical entropy of the scalar field on the background of
Reissner-Nordstrom black hole was calculated and the series
expansion of the entropy was derived near the black hole horizon.
Note that the higher-order terms of the entropy are divergent
[28]. Ref.[30] also discussed the statistical entropy of the
scalar field on the background of Reissner-Nordstrom black hole
and the series expansion of the entropy was derived near the black
hole horizon. Recently, Ref.[20] proposed the correction to state
density due to the generalized uncertainty principle is as
follows:

\begin{equation}
\label{eq3}
dn = \frac{d^3xd^3p}{(2\pi )^3}e^{ - \lambda p^2},
\end{equation}
where $p^2 = p^ip_i $, $\lambda $ is a constant characterized the
correction to Heisenberg uncertainty principle from the
gravitation. $\lambda $ is the same with Planck area
quantitatively.

In this paper, we calculate the statistical entropy of an
axisymmetric Kerr-Newman balck hole by using the correction
relation (\ref{eq3}) to state density due to the generalized
uncertainty principle. Since Kerr-Newman spacetime is
axisymmetric, in calculation, the integral interval is taken from
the black hole horizon surface to the hypersurface with the same
inherent radiation temperature measured by an observer at an
infinite distance.The inherent thickness of two surfaces are the
same with the minimal length quantitatively introduced in the
generalized uncertainty principle. And this thin layer clings to
the black hole horizon. Under the case without any artificial
cutoff and little mass approximation, we derive the series
expansions of the statistical entropies of Bose field and Fermi
field. And the leading terms of these expansions are directly
proportional to the area of the black hole horizon. Farther, this
series is convergent. that the black hole entropy can be expressed
as a convergence series. There does not exist any artificial
cutoff and little mass approximation. It is shown that there is
inherent relation between the black hole entropy and the horizon
area. Therefore, it makes people to have a better understanding of
the black hole statistical entropy in non-spherical symmetry
spacetimes. Because we adopt the quantum statistical method, the
difficulty of solving the wave equation in the conventional
brick-wall method is avoided. In the whole process, the
calculation is simple, the result is reasonable. We provide a
method for studying the quantum statistical entropy of a black
hole in a non-spherical symmetric spacetime. We take the simplest
function form of
temperature ($c = \hbar = G = K_B = 1)$.\vspace{0.5cm}\\
\textbf{\large 2. Entropy of Bose field}

Linear element of Kerr-Newman spacetime:

\[
ds^2 = - \left( {1 - \frac{2Mr - Q^2}{\rho ^2}} \right)dt^2 + \frac{\rho
^2}{\Delta }dr^2 + \rho ^2d\theta ^2
\]

\begin{equation}
\label{eq4} +\left[ {(r^2 + a^2)\sin ^2\theta + \frac{(2Mr -
Q^2)a^2\sin ^4\theta }{\rho ^2}} \right]d\varphi ^2-\frac{2(2Mr -
Q^2)a\sin ^2\theta }{\rho ^2}dtd\varphi,
\end{equation}
where $\rho ^2 = r^2 + a^2\cos ^\theta $, $\Delta = r^2 - 2Mr +
a^2 + Q^2$.

The radiation temperature of the black hole is as follows:

\begin{equation}
\label{eq5} T_ + = \frac{r_ + - r_ - }{4\pi (r_ + ^2 + a^2)},
\end{equation}
where $r_\pm = M\pm \sqrt {M^2 - a^2 - Q^2} $ are the locations of
outer and inner horizons of the black hole respectively. The area
of the black hole horizon is

\begin{equation}
\label{eq6} A(r_ + ) = 4\pi (r_ + ^2 + a^2).
\end{equation}
The natural radiation temperature got by the observer at rest at
an infinite distance is as follows [31]:

\begin{equation}
\label{eq7} T = \frac{T_ + }{\sqrt {- g_{tt}' } }.
\end{equation}
where $T_ + $ is the equilibrium temperature.

\begin{equation}
\label{eq8} g_{tt}' = \frac{g_{tt} g_{\varphi \varphi } -
g_{t\varphi }^2 }{g_{\varphi \varphi } }
 = - \frac{(r - r_ + )(r - r_ - )(r^2 + a^2\cos ^2\theta )}{(r^2 + a^2)^2 -
(r - r_ + )(r - r_ - )a^2\sin ^2\theta }.
\end{equation}
For Bose gas, the partition function $Z$ satisfies:

\begin{equation}
\label{eq9} \ln Z = - \sum\limits_i {g_i \ln (1 - e^{ - \beta
\varepsilon _i })} ,
\end{equation}
For spacetime (4), the area of two-dimensional curved surface at
arbitrary point $r$ outside horizon is

\begin{equation}
\label{eq10} A(r) = \int {dA = \int {\sqrt g d\theta d\varphi } }
,
\end{equation}
where $g = \left| {{\begin{array}{*{20}c}
 {g_{\theta \theta } } \hfill & {g_{\theta \varphi } } \hfill \\
 {g_{\varphi \theta } } \hfill & {g_{\varphi \varphi } } \hfill \\
\end{array} }} \right| = g_{\theta \theta } g_{\varphi \varphi }$. Then
the volume of the thin layer at arbitrary point $r$ outside the
horizon is as follows:

\begin{equation}
\label{eq11} dV = A(r)\sqrt {g_{rr} } dr.
\end{equation}
So, the partition function of the system at the thin layer with
arbitrary thickness at point $r$ outside the horizon is as
follows:

\[
\ln Z = - \int {A(r)\sqrt {g_{rr} } dr} \sum\limits_i {g_i } \ln
(1 - e^{ - \beta \varepsilon _i )}\] \[ = - \int {\frac{A(r)\sqrt
{g_{rr} } dr}{2\pi ^2}} \int\limits_0^\infty {p^2dpe^{ - \lambda
p^2}} \ln (1 - e^{ - \beta \omega _0 })
\]

\begin{equation}
\label{eq12}
 \approx \int {A(r)\sqrt {g_{rr} } dr} \int\limits_{m\sqrt { - g_{tt} '}
}^\infty {\frac{\beta _0 }{6\pi ^2(e^{\beta \omega _0 } - 1)}} p^3e^{ -
\lambda p^2}d\omega ,
\end{equation}
where $\beta = \beta _0 \sqrt { - g_{tt} '} $. The free energy of
the system is

\begin{equation}
\label{eq13} F = - \frac{1}{\beta _0 }\ln Z
 = - \int {A(r)\sqrt {g_{rr} } dr\int\limits_{m\sqrt { - g_{tt} '} }^\infty
{\frac{1}{6\pi ^2(e^{\beta \omega _0 } - 1)}} } p^3e^{ - \lambda p^2}d\omega
.
\end{equation}
The entropy of the system is

\[
S_B = \beta _0^2 \frac{\partial F}{\partial \beta _0 }
 = \beta _0^2 \int {A(r)\sqrt {g_{rr} } dr\int\limits_{m\sqrt { - g_{tt} '}
}^\infty {\frac{\omega e^{\beta \omega _0 }}{6\pi ^2(e^{\beta \omega _0 } -
1)^2}} } p^3e^{ - \lambda p^2}d\omega
\]

\[
 = \beta _0^2 \int \frac{A(r)\sqrt {g_{rr} }}{6\pi^2} dr\int\limits_{m\sqrt { - g_{tt} '}
}^\infty {\frac{\omega e^{\beta \textstyle{\omega \over {\sqrt { -
g_{tt} '} }}}}{(e^{\beta \textstyle{\omega \over {\sqrt { - g_{tt}
'} }}} - 1)^2}e^{ - \lambda \left( {\textstyle{{\omega ^2} \over {
- g_{tt} '}} - m^2} \right)}} (\frac{\omega ^2}{ - g_{tt} '} -
m^2)^{3 / 2}d\omega
\]

\begin{equation}
\label{eq14}
 = \frac{1}{6\pi ^2}\int {A(r)\sqrt {g_{rr} } dr\int\limits_{m\beta }^\infty
{\frac{xe^x}{(e^x - 1)^2}} } e^{ - \lambda \left( {\textstyle{{x^2} \over
{\beta ^2}} - m^2} \right)}(\frac{x^2}{\beta ^2} - m^2)^{3 / 2}dx.
\end{equation}
In the above calculation, we have used the relation among energy,
momentum and mass, $\textstyle{{\omega ^2} \over { - g_{tt} '}} =
p^2 + m^2$, and $m$ is a static mass of particle. In
Eq.(\ref{eq13}) the integral with respect to $r$ is near horizon,
so $g_{tt} '(r_ + ) \to 0$. Eq.(\ref{eq13}) can be reduced as:

\[
 S_B= \frac{1}{6\pi ^2}\int {A(r)\sqrt {g_{rr} } dr\int\limits_0^\infty
{\frac{x^4e^x}{\beta ^3(e^x - 1)^2}e^{ - \lambda \frac{x^2}{\beta ^2}}} }
dx
\]

\[
 = \frac{1}{6\pi ^2\beta _0^3 }\int {\frac{\sqrt {g_{\theta \theta }
g_{\varphi \varphi } g_{rr} } drd\theta d\varphi }{( - g_{tt} ')^{3 /
2}}\int\limits_0^\infty {\frac{x^4e^x}{(e^x - 1)^2}e^{ - \lambda
\frac{x^2}{\beta ^2}}} } dx
\]

\[
 = \frac{1}{6\pi ^2\beta _0^3 }\int\limits_0^\infty {\frac{dx}{4\sinh
^2\left( {\textstyle{x \over 2}} \right)}} \int {\frac{\sqrt {g_{\theta
\theta } g_{\varphi \varphi } g_{rr} } }{( - g_{tt} ')^{3 / 2}}x^4e^{ -
\lambda \frac{x^2}{\beta ^2}}} drd\theta d\varphi
\]

\begin{equation}
\label{eq15}
 = \frac{1}{6\pi ^2\beta _0^3 }\int\limits_0^\infty {\frac{dx}{4\sinh
^2\left( {\textstyle{x \over 2}} \right)}} I_1 (x,\varepsilon ),
\end{equation}

\noindent
where

\[
I_1 (x,\varepsilon ) = \int {\frac{\sqrt {g_{\theta \theta } g_{\varphi
\varphi } g_{rr} } }{( - g_{tt} ')^{3 / 2}}x^4e^{ - \lambda \frac{x^2}{\beta
^2}}} drd\theta d\varphi
\]

\[
 = \beta _0^4 \int {x^4\sqrt { - g_{\theta \theta } g_{\varphi \varphi }
g_{rr} g_{tt} '} \frac{\partial ^2}{\partial \lambda ^2}e^{ - \lambda
\frac{x^2}{\beta ^2}}} drd\theta d\varphi
\]

\begin{equation}
\label{eq16}
 = \beta _0^4 \int {(r^2 + a^2\cos ^2\theta )\sin \theta \frac{\partial
^2}{\partial \lambda ^2}} e^{ - \lambda \frac{x^2}{\beta ^2}}drd\theta
d\varphi .
\end{equation}
From (\ref{eq6}) and (\ref{eq7}), we obtain that outside the black
hole horizon at arbitrary point $R(R > r_ + )$ the natural
radiation temperature got by the observer at rest at an infinite
distance is different. It is related to angle $\theta $. When the
spacetime is spherically symmetric, outside the black hole horizon
at arbitrary point $R(R > r_ + )$ the natural radiation
temperature got by the observer at rest at an infinite distance is
the same. So, to calculate the statistical entropy of quantum
field near the black hole horizon, the integral interval usually
is taken as $[r_ + ,r_ + + \varepsilon ]$, and $\varepsilon $ is a
positive small quantity. From other viewpoint, the integral
interval is from the black hole horizon surface to the
hypersurface with the same inherent radiation temperature measured
by an observer at an infinite distance. It is shown that for
axisymmetric spacetime, to calculate the quantum statistical
entropy near outer of the black hole horizon, the integral
interval should be taken from the black hole horizon surface to
the hypersurface with the same inherent radiation temperature
measured by an observer at an infinite distance. From
Eq.(\ref{eq6}) and Eq.(\ref{eq7}), near outer of the horizon, when
$R$ satisfies

\begin{equation}
\label{eq17} R = r_ + + \frac{\varepsilon }{r_ + ^2 + a^2\cos
^2\theta } \quad ,
\end{equation}
an observer at rest at an infinite distance can get the
hypersurface with the same inherent radiation temperature near
outer of the black hole horizon. $r_ + $ is the location of the
black hole event horizon and satisfies $g_{tt} '(r_ + ) = 0$.
$\varepsilon $ is a positive small quantity. Near the horizon, the
metric can be simply written as $g_{tt} '(r) \approx (g_{tt}
')'(r_ + )(r - r_ + )$. From the metric (4), the minimal length is
obtained as

\begin{equation}
\label{eq18} \sqrt {\frac{e\lambda }{2}} = \int\limits_{r_ + }^R
{\sqrt {g_{rr} (r)} dr \approx \int\limits_{r_ + }^R {\frac{\sqrt
{r_ + ^2 + a^2\cos ^2\theta } }{\sqrt {(r - r_ + )(r_ + - r_ - )}
}} } dr = 2\sqrt {\frac{\varepsilon }{r_ + - r_ - }} ,
\end{equation}
Near the horizon, from Eq.(32) in Ref.[32], $I_1 (x,\varepsilon )$
can be expressed as:

\[
I_1 (x,\varepsilon ) = \beta _0^4 2\pi \int\limits_{r_ + }^R {(r^2 + a^2\cos
^2\theta )\sin \theta \frac{\partial ^2}{\partial \lambda ^2}e^{ - \lambda
\frac{x^2}{\beta ^2}}drd\theta }
\]

\[
 = \beta _0^4 2\pi \frac{\partial ^2}{\partial \lambda ^2}\int\limits_{r_ +
}^R \frac{(r^2 + a^2\cos ^2\theta )\sin \theta drd\theta }{\sum\limits_{n =
0} {\textstyle{1 \over {n!}}\left( {\textstyle{{\lambda x^2} \over { - \beta
_0^2 g_{tt} '(r)}}} \right)^n} }
\]

\begin{equation}
\label{eq19}
 \approx \beta _0^6 2\pi \frac{\partial ^2}{\partial \lambda
^2}\int\limits_{r_ + }^R { - \frac{(r_ + ^2 + a^2\cos ^2\theta )\sin \theta
d\theta }{\lambda x^2 - \beta _0^2 (g_{tt} ')'(r_ + )(r - r_ + ) + o((r - r_
+ )^2)}} (g_{_{tt} } ')'(r_ + )(r - r_ + )dr.
\end{equation}
Neglect the higher-order term, Eq.(\ref{eq18}) can be rewritten
as:

\[
I_1 (x,\varepsilon )\]
\[
 = \beta _0^4 2\pi \frac{\partial ^2}{\partial \lambda ^2}\int\limits_{r_ +
}^R {\left[ {1 - \frac{\lambda x^2}{\lambda x^2 - \beta _0^2 (g_{tt} ')'(r_
+ )(r - r_ + )}} \right]} (r_ + ^2 + a^2\cos ^2)\sin \theta d\theta dr
\]

\[
 = \beta _0^4 4\pi \frac{\partial ^2}{\partial \lambda ^2}\left[
{\varepsilon + \frac{\lambda x^2(r_ + ^2 + a^2)^2}{\beta _0^2 (r_ + - r_ -
)}\ln \left( {\frac{\lambda x^2}{\lambda x^2 + \beta _0^2 \frac{(r_ + - r_ -
)\varepsilon }{(r_ + ^2 + a^2)^2}}} \right)} \right]
\]

\[
 = \beta _0^6 4\pi \frac{(r_ + - r_ - )}{(r_ + ^2 +
a^2)^2}\frac{x^2\varepsilon ^2}{\lambda (\lambda x^2 + \beta _0^2 \frac{(r_
+ - r_ - )\varepsilon }{(r_ + ^2 + a^2)^2})^2}
\]

\begin{equation}
\label{eq20}
 = \frac{\beta _0^3 \pi ^3A(r_ + )}{\lambda }\frac{e^2x^2}{(x^2 + 2e\pi
^2)^2}.
\end{equation}
Substituting Eq. (\ref{eq19}) into Eq. (\ref{eq14}), we obtain

\begin{equation}
\label{eq21} S_B
 = \frac{\pi A(r_ + )e^2}{48\lambda }\int\limits_0^\infty {\frac{x^2}{\sinh
^2x}} \frac{dx}{(x^2 + e\pi ^2 / 2)^2}.
\end{equation}
Then, the integrand in Eq. (\ref{eq20}) can be regarded as a
complex function

\begin{equation}
\label{eq22} f(z) = \frac{z^2}{\sinh ^2z(z^2 + e\pi ^2 / 2)^2}
\end{equation}
When $n \ne 0$ are integers, $z = in\pi $ and $z = i\sqrt
{\textstyle{e \over 2}} \pi $ are two-order poles of $f(z)$. The
residues are as following respectively

\[
 - \frac{i}{2\sin ^2(\sqrt {e / 2} \pi )}\left[ {ctg(\sqrt {e / 2} \pi ) -
\frac{1}{2\sqrt {e / 2} \pi }} \right]
\]
and $ - 2in(n^2 + e / 2) / [\pi ^3(n^2 - e / 2)^3]$.

By the residue theorem, the entropy becomes

\[
S_B
 = \frac{\pi A(r_ + )e^2}{48\lambda }\left( \frac{\pi ctg(\sqrt {e / 2} \pi
)}{2\sin ^2(\sqrt {e / 2} \pi )} - \frac{\pi }{4\sqrt {e / 2} \sin
^2(\sqrt {e / 2} \pi )} \right.\]
\begin{equation}
\label{eq23} \left.+ \frac{2}{\pi ^2}\sum\limits_{n = 1}
{\frac{n(n^2 + e / 2)}{(n^2 - e / 2)^3}} \right).
\end{equation}
If the minimal length introduced in the generalized uncertainty
principle is assumed as

\begin{equation}
\label{eq24} \lambda = \frac{\pi ^2e^2}{48}\frac{2\sqrt {e / 2}
ctg(\sqrt {e / 2} \pi ) - 1}{\sqrt {e / 2} \sin ^2(\sqrt {e / 2}
\pi )}
\end{equation}
in the series expansion of the entropy Eq.(\ref{eq22}), the
leading term is a quarter of the horizon area, which satisfies B-H
entropy. Research on the correction to the black hole entropy is
one of the key issues. Many methods discussing the correction to
the black hole entropy have emerged [33-37], and the results are
valuable. However, we still need systematically discuss it. In
this paper, The correction to B-H entropy is obtained by
calculating the statistical entropies of Bose and Fermi field near
the horizon of the black hole. Note that the series expression of
the correction is convergent, so that our result is
reliable.\vspace{0.5cm}\\
\textbf{\large 3. Entropy of Fermi field}

For Fermi system, the partition function is

\begin{equation}
\label{eq25} \ln Z = \sum\limits_i {g_i \ln (1 + e^{ - \beta
\varepsilon _i })} .
\end{equation}
Based on Eq.(\ref{eq13}), we can obtain the entropy of Fermi
system.

\[
 = \frac{1}{6\pi ^2}\int {A(r)\sqrt {g_{rr} } dr\int\limits_0^\infty
{\frac{x^4e^x}{\beta ^3(e^x + 1)^2}e^{ - \lambda \frac{x^2}{\beta ^2}}} }
dx
\]

\[
 = \frac{1}{6\pi ^2\beta _0^3 }\int\limits_0^\infty {\frac{dx}{4\cosh
^2\left( {\textstyle{x \over 2}} \right)}} \int\limits_{r_ + }^{r_ + +
\varepsilon } {\frac{1}{g^2(r)}x^4e^{ - \lambda \frac{x^2}{\beta ^2}}}
A(r)dr
\]

\[
 = \frac{1}{6\pi ^2\beta _0^3 }\int\limits_0^\infty {\frac{dx}{4\cosh
^2\left( {\textstyle{x \over 2}} \right)}} I(x,\varepsilon )
\]

\begin{equation}
\label{eq26}
 = \frac{\pi A(r_ + )e^2}{48\lambda }\int\limits_0^\infty {\frac{x^2}{\cosh
^2x}} \frac{dx}{(x^2 + e\pi ^2 / 2)^2}.
\end{equation}
Therefore, we can obtain the entropy corresponding Fermi field
near the black hole horizon.

\[
S_F
 = \frac{\pi A(r_ + )e^2}{48\lambda }\left( \frac{\pi tg(\sqrt {e / 2} \pi
)}{2\cos ^2(\sqrt {e / 2} \pi )} - \frac{\pi }{4\sqrt {e / 2} \cos
^2(\sqrt {e / 2} \pi )} \right.
\]

\begin{equation}
\label{eq27} \left. + \frac{2}{\pi^2}\sum\limits_{n = 0} \frac{(n
+ 1/2)[(n + 1/2)^2 + e/2]}{[(n + 1 / 2)^2 - e / 2]^3} \right).
\end{equation}
If the minimal length introduced in the generalized uncertainty
principle is assumed as

\begin{equation}
\label{eq28} \lambda = \frac{\pi ^2e^2}{48}\frac{2\sqrt {e / 2}
tg(\sqrt {e / 2} \pi ) - 1}{\sqrt {e / 2} \cos ^2(\sqrt {e / 2}
\pi )}
\end{equation}

\noindent the leading term in series expansion of Fermi entropy is
directly proportional to the horizon area, which satisfies the
Bekenstein-Hawking
formula.\vspace{0.5cm}\\
\textbf{\large 4. Conclusion }

Taking into account the effect of the generalized uncertainty
principle on the state density, we calculate the statistical
entropy of the Bose and Fermi field near the black hole horizon.
The series expansion of the statistical entropy is obtained
without any artificial cutoff and little mass approximation. When
dimensionless constant $\lambda $ in the generalized uncertainty
principle satisfies Eq.(\ref{eq23}) and Eq.(\ref{eq27}), the
leading term in the series expression of the statistical entropy
is proportional to the area of horizon.It satisfies the conditions
of B-H entropy In our calculation, the difficulty in solving the
wave equation is overcome by the quantum statistical method. In
our calculation, the difficulty in solving the wave equation is
overcome by the quantum statistical method. For spherically
symmetric spacetime, to calculate the black hole entropy we
proposed the integral interval should be: from the black hole
horizon surface to the hypersurface with the same inherent
radiation temperature measured by an observer at an infinite
distance near outer of the black hole horizon. It also provides a
way for studying the quantum statistical entropy of a
black hole in a non-spherical symmetric spacetime.\\
\textbf{ACKNOWLEDGMENT}

This project was supported by the Shanxi Natural Science
Foundation of China under Grant No.
2006011012.\vspace{0.5cm}\\
\textbf{Reference}

[1] J. D. Bekenstein, Phys . Rev. D \textbf{7}, 2333 (1973).

[2] S. W. Hawking, Commun. Math. Phys. \textbf{43}, 199 (1975).

[3] J. M. Bardeen, B. Carter, S. W. Hawking, Math. Phys. \textbf{31}, 161
(1973).

[4] D. Hochberg, T.W. Kephart, J.W. York, Phys. Rev. D
\textbf{48}, 479 (1993).

[5] T. Padmanaban, Phys. Lett.s A. \textbf{136}, 203 (1989).

[6] H. Lee, S.W. Kim and W.T. Kim, Phys. Rev. D \textbf{54}, 6559
(1996)

[7] G't Hooft.  Nucl. Phys. B \textbf{256}, 727 (1985).

[8] G. Cognola and P. Lecca, Phys. Rev. D \textbf57, 1108 (1998).

[9] R.G. Cai, J.Y. Ji and K.S. Soh, Class. Quantum. Grav. \textbf{15}, 2783
(1998).

[10] M.H. Lee and J. K. Kim, Phys. Rev. D \textbf{54}, 3904
(1996).

[11] A. Ghosh and P. Mitra, Phys. Rev. Lett. \textbf{73}, 2521 (1994).

[12] J. Jing and M. L. Yan, Phys. Rev. D \textbf{60}, 084015
(1999).

[13] M. Kenmoku, K. Ishimoto, K. K. Nandi and K. Shigemoto, Phys.
Rev. D \textbf{73}, 064004 (2006).

[14] R. Zhao, J.F. Zhang and L.C. Zhang, Nucl Phys. B
\textbf{609}, 247 (2001).

[15] Xiang, L. and Zheng, Z. Phys. Rev. D \textbf{62}, 104001
(2000).

[16] F. He and Z. Zhao and S. W. Kim, Phys. Rev. D \textbf{64},
044025 (2001).

[17] C. J. Gao and Y. G. Shen, Phys. Rev. D \textbf{65}, 084043
(2002).

[18] A. Kempf, G. Mangano and R. B. Mann, Phys. Rev. D
\textbf{52}, 1108 (1995).

[19] L. N. Chang, D. Minic, N. Okamura and T. Takeuchi, Phys. Rev.
D \textbf{65}, 125028 (2002).

[20] K. Nouicer, Phys. Lett. B \textbf{646}, 63 (2007).

[21] M. R. Setare, Phys. Rev. D \textbf{70}, 087501 (2004).

[22] A. J. M. Medved and E. C. Vagenas, Phys. Rev. D \textbf{70},
124021 (2004).

[23] M. R. Setare, Int. J. Mod. Phys. A \textbf{21}, 1325 (2006).

[24] X. Li, Phys. Lett. B \textbf{540}, 9 (2002).

[25] R. Zhao, Y. Q. Wu and L. C. Zhang, Class. Quantum. Grav. \textbf{20},
4885 (2003).

[26] R. Zhao and S. L. Zhang, Gen. Rel. Grav. \textbf{36}, 2123
(2004).

[27] R. Zhao and S. L, Zhang, Gen. Rel. Grav \textbf{36}, 2539 (2004).

[28] W. Kim, Y. W. Kim and Y. J. Park, Phys. Rev. D \textbf{74},
104001 (2006).

[29] W. Kim, Y. W. Kim and Y. J. Park, Phys. Rev. D \textbf{75},
127501 (2007).

[30] M. Yoon, J. Ha and W. Kim, Phys. Rev. D \textbf{76}, 047501
(2007).

[31] R. C. Tolman, \textit{Relativity, Thermodynamics and Cosmology.} (Oxford: Oxford University Press 1934)

[32] Y. W. Kim, and Y. J. Park, Phys. Lett. B \textbf{655}, 172
(2007).

[33] A. Chatterjee and P. Majumdar, Phys. Rev. D\textbf{71}, 024003 (2005).

[34] Y. S. Myung, Phys. Lett. B\textbf{579}, 205 (2004).

[35] R. Zhao and S. L. Zhang, Phys. Lett. B \textbf{641}, 208
(2006).

[36] R. Zhao and S. L. Zhang, Phys. Lett. B \textbf{641,} 318
(2006).

[37] R. Zhao, H. X. Zhao and S. Q. Hu,. Mod. Phys. Lett. A, \textbf{22},
1737 (2007).

\end{document}